\documentclass{article}
\usepackage{spconf,graphicx}
\usepackage{amsmath,amssymb,fixmath}
\usepackage{booktabs}
\usepackage{cite}
\usepackage{url,hyperref}
\usepackage{color}
\usepackage[utf8]{inputenc} 

\usepackage{multirow}
\usepackage{makecell}
\usepackage{multicol}
\usepackage{tabularx}
\usepackage{color}

\usepackage[caption=false]{subfig}

\usepackage{float}
\usepackage[dvips]{epsfig}
\usepackage{multirow}

\DeclareMathOperator{\E}{\mathbb{E}}
\DeclareMathOperator{\bx}{\mathbold{x}}
\DeclareMathOperator{\bz}{\mathbold{z}}
\DeclareMathOperator{\bh}{\mathbold{h}}
\DeclareMathOperator{\bxhat}{\mathbold{\hat{x}}}

\newcommand{\rrredit}[1]{\textcolor{black}{#1}}
\newcommand{\eeeeedit}[1]{\textcolor{black}{#1}}
\newcommand{\mmmmmedit}[1]{\textcolor{black}{#1}}

\newcommand{\redit}[1]{\textcolor{black}{#1}}
\newcommand{\rredit}[1]{\textcolor{black}{#1}}
\newcommand{\eedit}[1]{\textcolor{black}{#1}}
\newcommand{\eeedit}[1]{\textcolor{black}{#1}}
\newcommand{\eeeedit}[1]{\textcolor{black}{#1}}
\newcommand{\medit}[1]{\textcolor{black}{#1}}

\newcommand{\mosfa}{$4.20$}
\newcommand{\mosfb}{$4.18$}
\newcommand{\mosma}{$4.21$}
\newcommand{\mosmb}{$4.31$ } 

\title{Parallel waveform synthesis based on generative adversarial networks with voicing-aware conditional discriminators}

\name{Ryuichi Yamamoto$^{1}$, Eunwoo Song$^{2}$, Min-Jae Hwang$^{3}$ and Jae-Min Kim$^{2}$}
\address{
$^1$LINE Corp., Tokyo, Japan\\ 
$^2$NAVER Corp., Seongnam, Korea\\
$^3$Search Solutions Inc., Seongnam, Korea
}

\begin{document}
\fontsize{9.3}{11.3}\selectfont
\maketitle
\begin{abstract}
    This paper proposes voicing-aware conditional discriminators for Parallel WaveGAN-based waveform synthesis systems. In this framework, we \eeedit{adopt a projection-based conditioning method that can significantly improve the discriminator's performance}. Furthermore, \eeedit{the conventional discriminator is separated into} two waveform discriminators \redit{for \rredit{modeling} voiced and unvoiced speech}. \rredit{As} each discriminator learns the distinctive characteristics of the harmonic and noise components, respectively, the adversarial training process becomes more efficient, \redit{allowing the generator to produce more realistic speech waveforms}. Subjective test results demonstrate the superiority of the proposed method over the conventional \rredit{Parallel WaveGAN and  WaveNet systems}. In particular, our \redit{speaker-independently trained} model within a FastSpeech 2 based text-to-speech framework achieves the mean opinion scores of \mosfa, \mosfb, \mosma, and \mosmb for four Japanese speakers, respectively.
\end{abstract}

\section{Introduction}
\label{sec:introduction}

    Deep generative models in text-to-speech (TTS) frameworks have significantly improved the perceptual quality of synthetic speech signals~\cite{ze2013statistical,Wang2018WNComparison}.
    In particular, \medit{the} autoregressive generative models, such as WaveNet, have shown superior quality over conventional parametric vocoders~\cite{oord2016wavenet, kalchbrenner2018efficient, tamamori2017speaker, hayashi2017multi, song2019excitnet}. 
    However, they suffer from slow generation due to their autoregressive nature and thus are limited in their applications to real-time scenarios.

    
    \medit{To achieve real-time TTS systems,} non-autoregressive \medit{waveform} synthesis models have been proposed based on teacher-student frameworks~\cite{Oord2018ParallelWF, Ping2018ClariNetPW}, normalizing flows~\cite{prenger2018waveglow,kim2018flowavenet}, or generative adversarial networks (GANs)~\cite{Yamamoto2019, kumar2019melgan}.
    \medit{Specifically, i}n \eedit{our previous work\redit{, we} proposed} the \textit{Parallel WaveGAN} \rrredit{methods} \eeeeedit{~\cite{yamamoto2020parallel, song2021improved}}, characterized by efficient training and fast inference while maintaining a quality that is competitive to the state-of-the-art Parallel WaveNet.
    \medit{
        However, as it is insufficient for a single discriminator to distinguish the complex nature of speech signal --- e.g.,  voiced and unvoiced characteristics --- the generated speech often suffers from unnatural artifacts.
        \mmmmmedit{
            In addition, this problem becomes more severe when the training database has more \redit{diversity} such as in the scenario of speaker-independent modeling.
        }
    }
    
    
    To address the aforementioned problems, we propose \eedit{voicing-aware conditional} discriminators for Parallel WaveGAN. 
    In this method, we adopt a projection-based conditioning framework that incorporates acoustic features into the discriminators~\cite{miyato2018cgans}.
    This enables the discriminator to classify the input speech well to be consistent with the given acoustic features.
    Furthermore, we introduce two separate voicing-aware discriminators that individually \rredit{model the voiced and unvoiced speech}, respectively.
    In detail, one discriminator is designed to have long receptive fields for capturing \medit{\eeedit{slowly varying harmonic components}, which mainly represents the \eeedit{voiced speech}}; whereas the other has small receptive fields for capturing \medit{\eeedit{rapidly varying noise components}} \redit{of unvoiced speech}.
    Because each discriminator learns the distinctive characteristics of the \rredit{harmonic and noise components}, respectively, the adversarial training process becomes more effective.
    
    We investigate the performance of our proposed method by conducting perceptual listening tests in a TTS framework.
    Specifically, a speaker-independently trained Parallel WaveGAN with the FastSpeech 2 acoustic model significantly outperforms \redit{the conventional Parallel WaveGAN and similarly configured WaveNet systems}, achieving mean opinion scores of \mosfa, \mosfb, \mosma, and \mosmb for four Japanese speakers, respectively.
    
\section{Related work}
\label{sec:related work}

    
    There have been several attempts to improve \eedit{the discriminator's performance for} GAN-based neural waveform synthesis 
    \eedit{systems}. 
    \eedit{For instance,} MelGAN\eedit{\cite{kumar2019melgan}} and VocGAN\eedit{\cite{yang2020vocgan}} employ multi-scale discriminators to learn the waveform structure on different time scales.
    \redit{GAN-TTS\eedit{\cite{binkowski2019high}} adopts a blend of multiple conditional and unconditional discriminators based on multi-frequency random windows.}
    Although \redit{these multi-resolution architectures} are found to be effective for high perceptual quality, their methods tend to require complicated discriminators (e.g., hierarchically-nested \redit{joint conditional and unconditional} discriminators~\cite{yang2020vocgan}), which are more difficult to train. 
    To keep the discriminator simple yet effective, our proposed method adopts two separate discriminators, which can explicitly focus on the distinctive \rredit{voiced and unvoiced} characteristics of speech.

\section{Method}
\label{sec:method}

\subsection{Parallel WaveGAN}
\label{subsec:pwg}

    \eedit{Parallel WaveGAN is a non-autoregressive WaveNet model that generates} a time-domain speech waveform from the corresponding conditional acoustic parameters~\cite{yamamoto2020parallel}.
    Specifically, the conventional Parallel WaveGAN \eedit{consists of} a non-causal WaveNet generator, $G$, and a single \redit{convolutional neural network (CNN)} discriminator, $D$.
    Based on GANs~\cite{goodfellow2014generative}, the generator learns a distribution of realistic waveforms by trying to deceive the discriminator into recognizing the generated samples as \textit{real}.
    Moreover, the discriminator is trained to correctly classify the generated sample as \textit{fake} while classifying the ground truth as \textit{real}.
    By \eedit{combining adversarial training with an} \redit{auxiliary} multi-resolution short-time Fourier transform (STFT) loss function, Parallel WaveGAN learns the time-frequency characteristics of realistic speech efficiently.

\subsection{Proposed Parallel WaveGAN with voicing-aware conditional discriminators}
\label{subsec:vuvd}

	\begin{figure}[!t]
	\begin{minipage}[t]{.49\linewidth}
	\centerline{\epsfig{figure=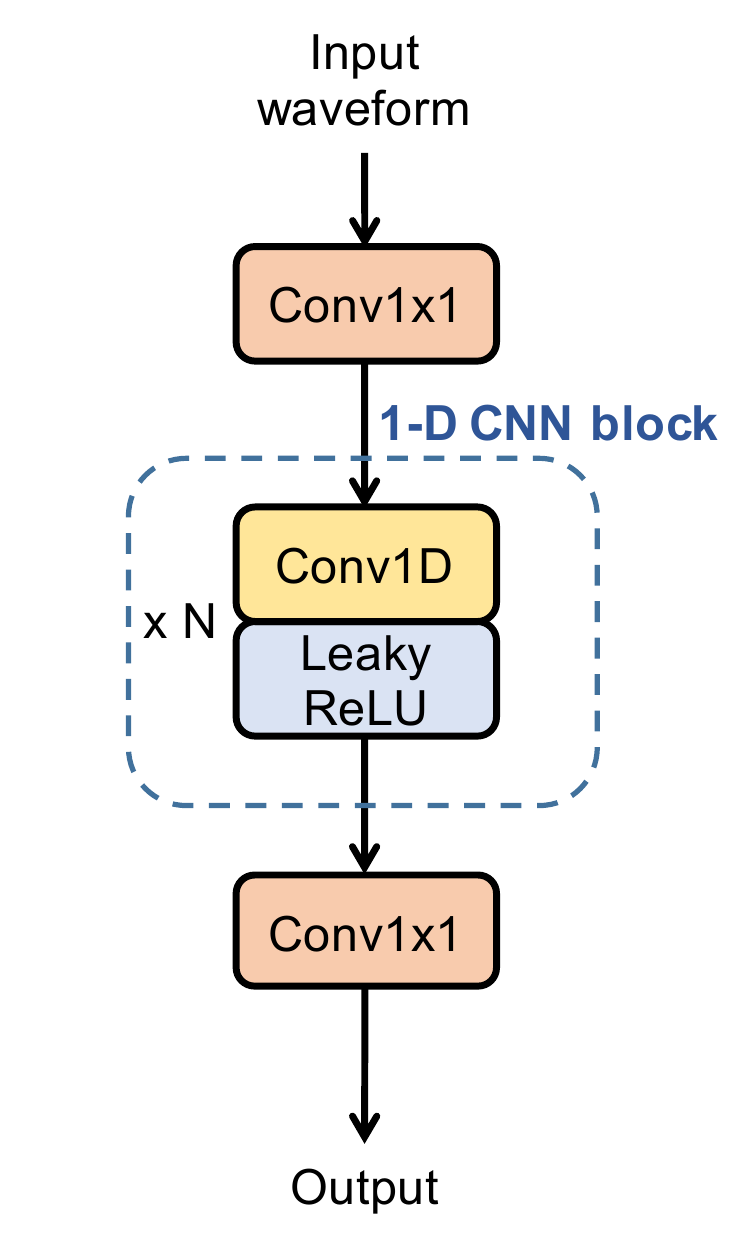,width=33.1mm}}
	\vspace*{-1pt}
	\centerline{(a)}  \medskip
	\end{minipage}
	\begin{minipage}[t]{.49\linewidth}
	\centerline{\epsfig{figure=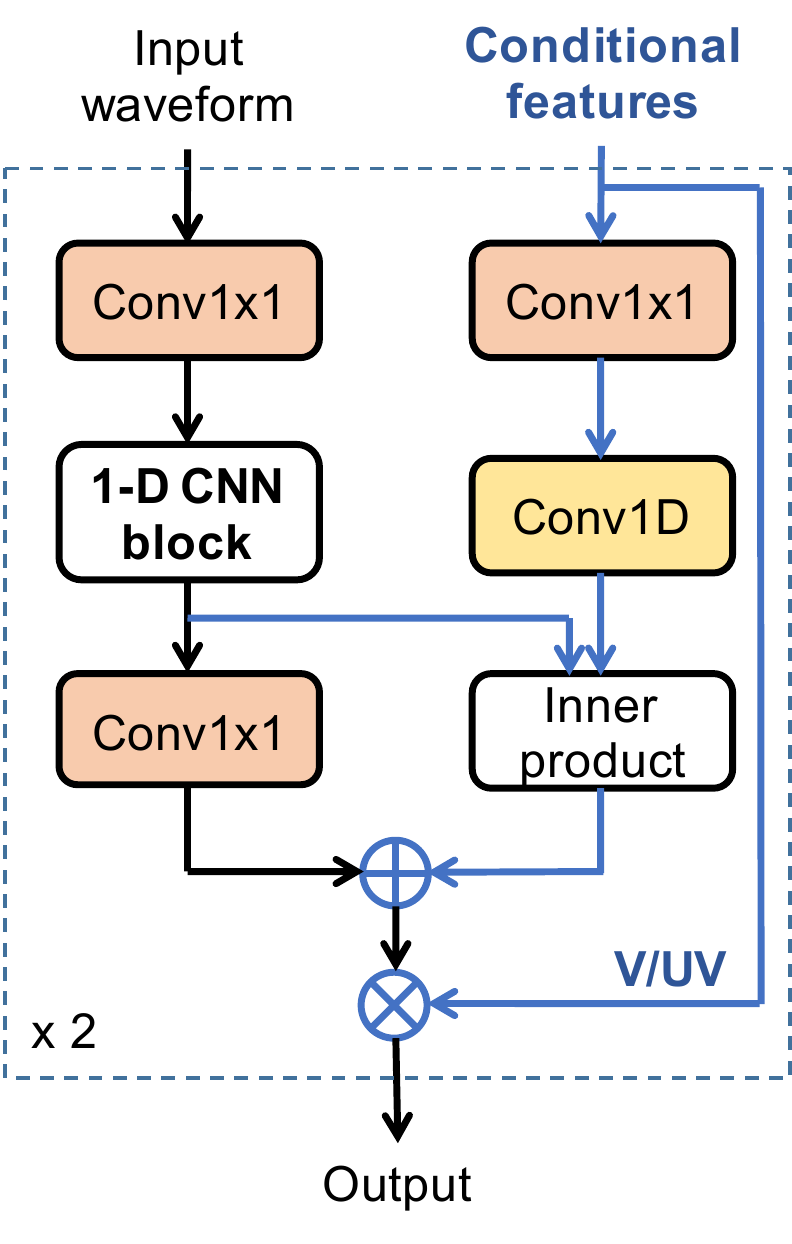,width=35mm}}
	\vspace*{-1pt}	
	\centerline{(b)}  \medskip
	\end{minipage}	
	\caption{\small Block diagram of (a) a conventional and (b) the proposed discriminators. Note that in the proposed method, two separate discriminators with different dilation factors of the 1-D convolutional neural network (CNN) blocks were used for \eeeedit{modeling} the voiced and unvoiced segments, respectively.}
	\vspace*{-1mm}
	\label{fig:vuvd}
	\end{figure}

    Fig.~\ref{fig:vuvd} \eedit{depicts an} overview of the \eedit{Parallel WaveGAN's discriminator}.
    Compared with the conventional method (Fig.~\ref{fig:vuvd}(a)), there are two main improvements in the proposed method (Fig.~\ref{fig:vuvd}(b)), as follows. First, we adopt \eedit{a} projection-based \eedit{conditioning method where the} acoustic features \eedit{are incorporated} into the discriminator as conditional \eedit{inputs}~\cite{miyato2018cgans}.
    This helps the discriminator to better classify the input signals to be consistent with the given conditional features.
    Second, we replace the conventional discriminator with voicing-aware ones using a \rredit{voiced and unvoiced} binary flag (V/UV).
    Considering the fact that the voiced and unvoiced segments of speech signals have distinctive characteristics, the two separate discriminators independently operate to capture each segment, respectively. Note that voicing masks are used to make each discriminator see only the region of its interest.    
    
    The voiced segment can be characterized by slowly evolving harmonic components.
    To control these components, we design the first discriminator with a dilated CNN~\cite{oord2016wavenet}.
    Note that the use of dilated convolution allows the discriminator to increase the size of the receptive field while keeping a small number of parameters.
    With a sufficient size of the receptive field, the discriminator not only covers long-term variations of the harmonic component, but also penalizes any unwanted aperiodic noise components in the voiced regions.
    On the other hand, the second discriminator is composed of a non-dilated CNN with a small receptive field (i.e., a dilation factor of 1 in the 1-D CNN block).
    Because the characteristics of the noise component vary rapidly, employing a short window is advantageous to focus on the detailed high-frequency structure of speech.    

\subsection{Training objectives}
\label{subsec:objective}
    To train the proposed models, we adopt the least-squares GANs thanks to their training stability~\cite{mao2017least}. 
    The training objectives for the discriminators and the generator are defined as follows:
    \begin{align}
        \min_{D} & \E_{\bz, \bh}[(1 - D(\bx, \bh))^2] \nonumber \\
        &+ \E_{\bz, \bh}\left[D(G(\bz, \bh), \bh)^2\right], \forall D \in \{D^{\mathrm{v}}, D^{\mathrm{uv}}\} \label{eq:dloss} \\
     \min_{G} & \E_{\bx, \bz, \bh}\left[ L_{\mathrm{mr\_stft}}(\bx, G(\bz, \bh)) \right] \nonumber \\
        &+ \frac{1}{2} \lambda_{\mathrm{adv}} \E_{\bz, \bh}\left[\sum_{D \in \{D^{\mathrm{v}}, D^{\mathrm{uv}}\}} (1 - D(G(\bz, \bh), \bh))^2\right],
    \label{eq:gloss}
    \end{align}    
    where $\bz$, $\bh$, and $\bx$ denote the Gaussian noise, conditional acoustic \eedit{features}, and the target speech waveform, respectively; $D^{\mathrm{v}}$ and $D^{\mathrm{uv}}$ are the voiced \eedit{and} unvoiced discriminators, respectively; and $\lambda_{\mathrm{adv}}$ represents a hyperparameter that balances the two adversarial \eedit{losses} and the multi-resolution STFT loss \eedit{defined as follows:}
    \begin{equation}
        L_{\mathrm{mr\_stft}}(\bx, \bxhat) = \frac{1}{M}\sum_{m=1}^{M}L^{(m)}_{\mathrm{stft}}(\bx, \bxhat), \label{eq:multispec}
    \end{equation}
    where \rredit{$\bxhat$ represents the generated waveform; and} $L^{(m)}_{\mathrm{stft}}(\bx, \bxhat)$ denotes the $m^{th}$ STFT loss, represented by the sum of \textit{spectral convergence} and \textit{log STFT magnitude losses}\footnote{
    \rredit{The detailed setups for designing the multi-resolution STFT loss are the same as those in the original Parallel WaveGAN~\cite{yamamoto2020parallel}}.
    }~\cite{arik2019fast}.
    \redit{Note that unlike the original Parallel WaveGAN~\cite{yamamoto2020parallel}, the discriminator is now divided into distinctive voiced and unvoiced parts, and the generator is designed to deceive both of them.}

\section{Experiments}

\label{sec:experiments}

\subsection{Experimental setup}

    \subsubsection{Data and feature configurations}

    The experiments used four phonetically and prosodically rich speech corpora recorded by two female (F1, F2) and two male (M1, M2) Japanese professional speakers.
    The speech signals were sampled at 24 kHz, and each sample was quantized by 16 bits.
    Each corpus included 5,000 utterances, among which 4,500, 250, and 250 samples were used for training, validation, and evaluation, respectively. The training data size for each speaker was between 5.5 and 5.9 hours.

    The acoustic features were extracted using an improved time-frequency trajectory excitation vocoder at the analysis intervals of 5 ms~\cite{song2017effective} and included 40-dimensional line spectral frequencies, the fundamental frequency, the energy, the \eeeedit{binary} \rredit{V/UV flag}, a 32-dimensional slowly evolving waveform, and a 4-dimensional rapidly evolving waveform, all of which constituted a 79-dimensional feature vector\footnote{\redit{We have also tried mel-spectrograms as acoustic features but found that the vocoder parameters were more effective to avoid buzzy synthetic speech.}}.
    The acoustic features were then normalized to have zero mean and unit variance using the statistics of the training data.
    
	\subsubsection{Model details}
	
    \begin{table}[t]
      \caption{\small The dilation factors and receptive fields in the 1-D CNN blocks of the voicing-aware discriminators.}
      \vspace{1mm}
      \label{tab:vuvd}
      \centering
      \scalebox{0.90}{
      {\renewcommand\arraystretch{0.90}
      \begin{tabular}{cccc}
        \toprule
        \textbf{Discriminator} & \textbf{Dilation factors} & \textbf{Receptive field}\\
        \midrule
        $D^{\mathrm{v}}$ & [1, 2, 4, 8, 16, 32] & 127 \\
        $D^{\mathrm{uv}}$ & [1, 1, 1, 1, 1, 1] & 13 \\
        \bottomrule
      \vspace{-4mm}
      \end{tabular}
      }
      }
    \end{table}

    The proposed Parallel WaveGAN consists of a WaveNet-based generator and voiced and unvoiced discriminator\rredit{s}. The generator comprises 30 layers of dilated residual 1-D convolution blocks with three exponentially increasing dilation cycles~\cite{yamamoto2020parallel}.
    The number of residual and skip channels was set to 64, and the convolution filter size was set to 5.
    The size of the receptive field for the generator was 12,277. The discriminators for the voiced and unvoiced regions were each composed of a 1-D CNN block and 1-by-1 convolution layers.
    Each 1-D CNN block contains six convolution layers interleaved with leaky ReLU activation.
    The number of channels and kernel size in the 1-D CNN blocks were set to 64 and 3, respectively.
    The dilation factors and receptive fields\footnote{The receptive field for the discriminator of the voiced regions was kept not too large because a larger receptive field poses training difficulty.} of the 1-D CNN blocks for the voicing-aware discriminator are summarized in Table~\ref{tab:vuvd}.
    For conditional input, a 1-D convolution with the kernel size of the discriminator’s receptive field was used before the inner product projection. 
    The number of channels was 64, the same as in the 1-D CNN block.

    At the training stage, the multi-resolution STFT loss was computed by the sum of three different STFT losses, as described in Parallel WaveGAN~\cite{yamamoto2020parallel}.
    The hyperparameter $\lambda_{\mathrm{adv}}$ in equation~(\ref{eq:gloss}) was chosen to be 4.0.
    The models were trained for 400K steps with a RAdam optimizer with $\beta_{1}=0.9$, $\beta_{2}=0.999$, and $\epsilon=1e^{-6}$~\cite{liu2020radam}.
    The discriminators were fixed for the first 100K steps, and the models were jointly trained afterwards.
    The mini-batch size was set to eight, and the length of each audio clip was set to 24K time samples (1.0 second).
    The initial learning rate was set to $0.0001$ for both the generator and discriminators.
    The learning rate was reduced by half for every 200K steps.
    \rrredit{We trained Parallel WaveGAN models with two NVIDIA Tesla V100 GPUs, which took about 44 and 58 hours for conventional and proposed Parallel WaveGAN systems, respectively.}

    To validate our discriminator design choices, we investigated six Parallel WaveGAN systems with different discriminator configurations, as described in Table~\ref{tab:mos_anasyn}. Note that all the Parallel WaveGAN systems used the same generator architecture and training configurations \redit{as described above}; they only differed in the discriminator settings. 
    \rrredit{Therefore, the proposed method retained the original Parallel WaveGAN’s fast inference speed.}

    As a baseline system, we used the autoregressive Gaussian WaveNet~\cite{Ping2018ClariNetPW}, which consists of 24 layers of dilated residual convolution blocks with \redit{exponentially increasing} four dilation cycles.
    The number of residual and skip channels was set to 128, and the filter size was set to 3.
    The model was trained for 1M steps with a RAdam optimizer. 
    The learning rate was set to $0.001$ and reduced by half for every 200K steps.
    The mini-batch size was set to eight, and the length of each audio clip was set to 12 K time samples (0.5 seconds).
    The log-scale parameters of Gaussian were clipped at \rredit{$-9.0$} during training to reduce noisy artifacts~\cite{Ping2018ClariNetPW}.

    Across all the neural vocoders, the input auxiliary features were up-sampled by nearest neighbor interpolation followed by 1-D convolutions so that the time resolution of the auxiliary features matched the sampling rate of the speech waveforms~\cite{odena2016deconvolution,Yamamoto2019}.
    For the models using conditional discriminators, the up-sampled features were used as the conditional input.
    Note that the binary V/UV flag used in the voicing-aware discriminator was up-sampled to the sample level by repetition as an exception.
    All the vocoder models were trained in a speaker-independent manner by putting all the speaker’s data together. 

    \subsection{Evaluation}
    \label{sec:evaluation}

	\begin{table*}[!t]   
	\begin{center}         
	\caption{\small \eeeeedit{MOS} test results with 95\% confidence intervals in \rredit{analysis/synthesis}: \eeeeedit{The} speech samples were generated using the acoustic features extracted from the recorded speech. 
	PWG denotes Parallel WaveGAN for short. Note that systems S2 and S3 used $D^{\mathrm{v}}$ as the primary discriminator. All the models were trained in a speaker-independent manner.}
    \vspace{1mm}
	\label{tab:mos_anasyn}
	\scalebox{0.99}{
	{\small        
	\begin{tabular}{cllllcccc}
	\Xhline{2\arrayrulewidth}
    \multirow{2}*{System} & \multirow{2}*{{Model}} & Voiced & {Unvoiced} & {Discriminator} & \multicolumn{4}{c}{MOS} \\
     & & \redit{segments} & \redit{segments} & {conditioning} & {F1} & {F2} & {M1} & M2 \\
    \hline
    S1 & WaveNet & - & - & -                  & 3.64$\pm$0.12  & 3.83$\pm$0.11  & 3.33$\pm$0.12  & 3.13$\pm$0.11 \\
    S2 & PWG & - & - & -                      & 3.61$\pm$0.11  & 3.55$\pm$0.11  & 3.57$\pm$0.12  & 3.61$\pm$0.11 \\
    S3 & PWG-cGAN-D & - & - & Yes             & 4.04$\pm$0.10  & 3.95$\pm$0.10  & 3.91$\pm$0.11  & 3.97$\pm$0.10  \\
    S4 & PWG-V/UV-D & $D^{\mathrm{v}}$ & $D^{\mathrm{v}}$ & Yes & 3.60$\pm$0.12  & 3.59$\pm$0.11  & 3.34$\pm$0.11  & 3.48$\pm$0.11  \\
    S5 & PWG-V/UV-D & $D^{\mathrm{uv}}$ & $D^{\mathrm{v}}$ & Yes & 3.67$\pm$0.11  & 3.48$\pm$0.11  & 3.29$\pm$0.12  & 3.38$\pm$0.11 \\
    S6 & PWG-V/UV-D & $D^{\mathrm{uv}}$ & $D^{\mathrm{uv}}$ & Yes & 3.77$\pm$0.11  & 3.88$\pm$0.10  & 3.57$\pm$0.11  & 3.34$\pm$0.11 \\
    \textbf{S7} & \textbf{PWG-V/UV-D (proposed)} & $D^{\mathrm{v}}$ & $D^{\mathrm{uv}}$ & Yes & \textbf{4.11$\pm$0.10} & \textbf{4.05$\pm$0.10}  & \textbf{4.04$\pm$0.10} & \textbf{4.08$\pm$0.10} \\
    \bottomrule
    R1 & Recordings & - & - & -                 & 4.63$\pm$0.08  & 4.67$\pm$0.07  & 4.61$\pm$0.08  & 4.64$\pm$0.08 \\
    \Xhline{2\arrayrulewidth}
	\end{tabular}}    
	}
	\end{center}
    \vspace{-2mm}
	\end{table*}

	\begin{figure}[!t]
	\begin{minipage}[t]{.32\linewidth}
	\centerline{\epsfig{figure=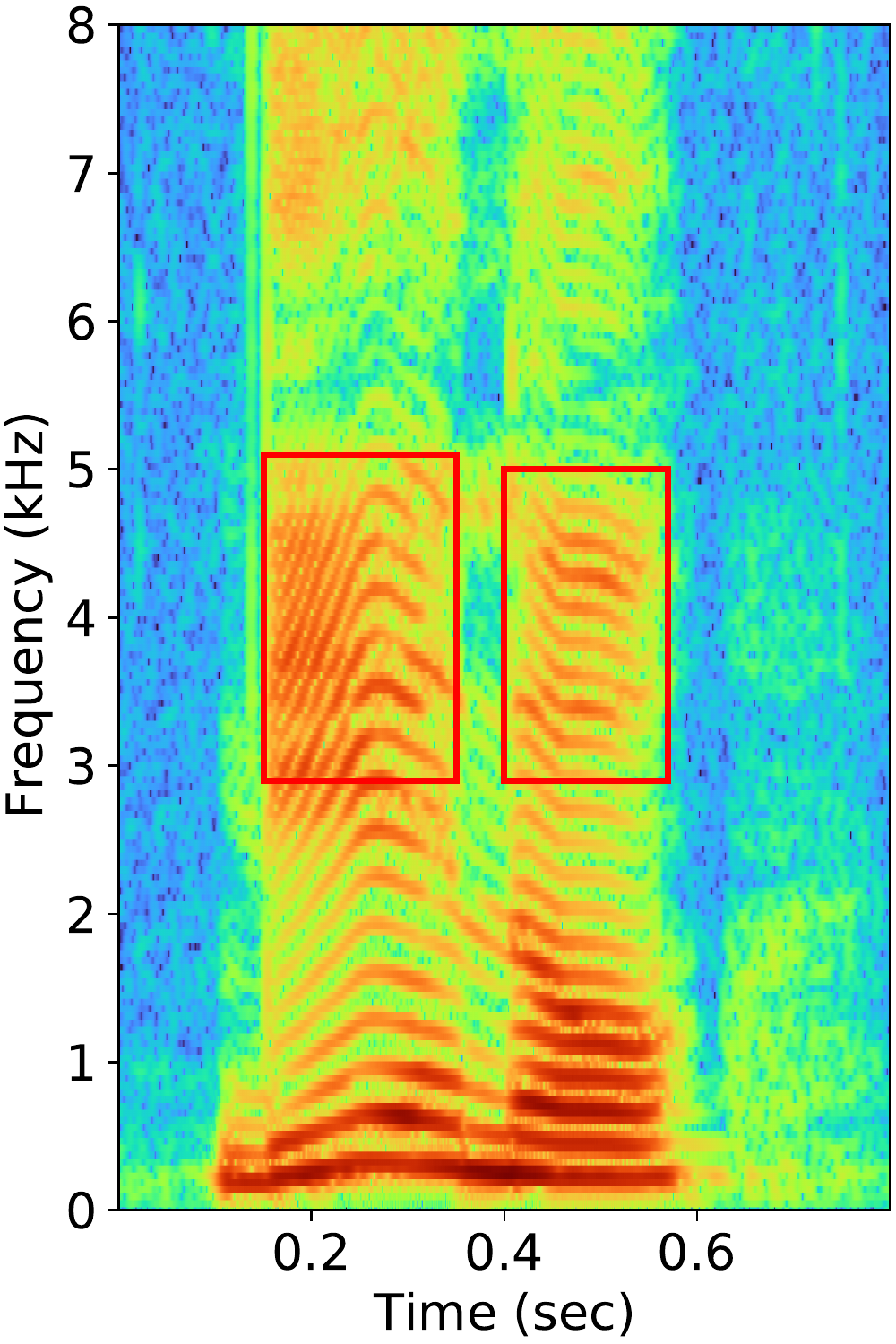,width=28mm}}
	\centerline{(a)}  \medskip
	\end{minipage}
	\begin{minipage}[t]{.32\linewidth}
	\centerline{\epsfig{figure=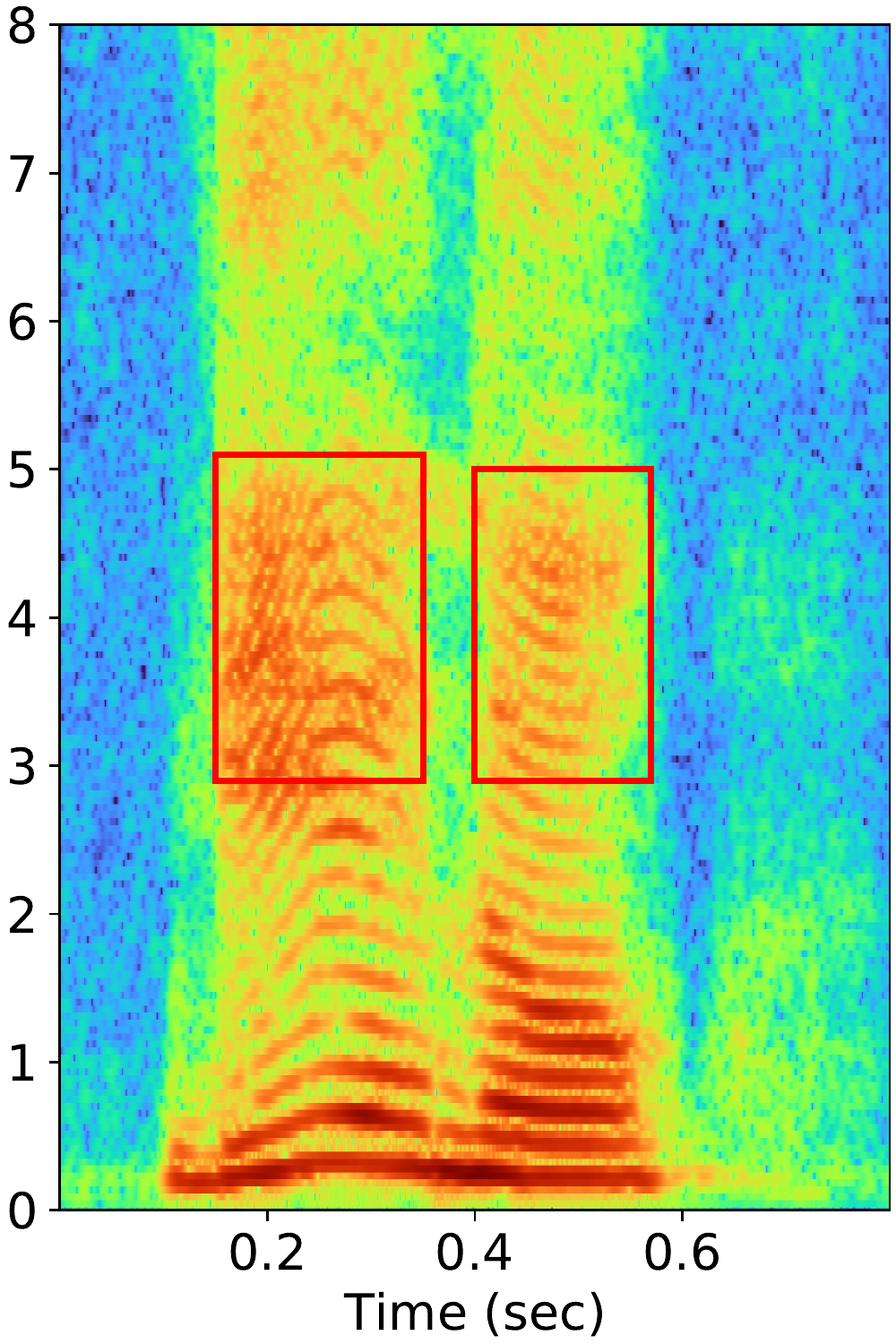,width=28mm}}
	\centerline{(b)}  \medskip
	\end{minipage}	
	\begin{minipage}[t]{.32\linewidth}
	\centerline{\epsfig{figure=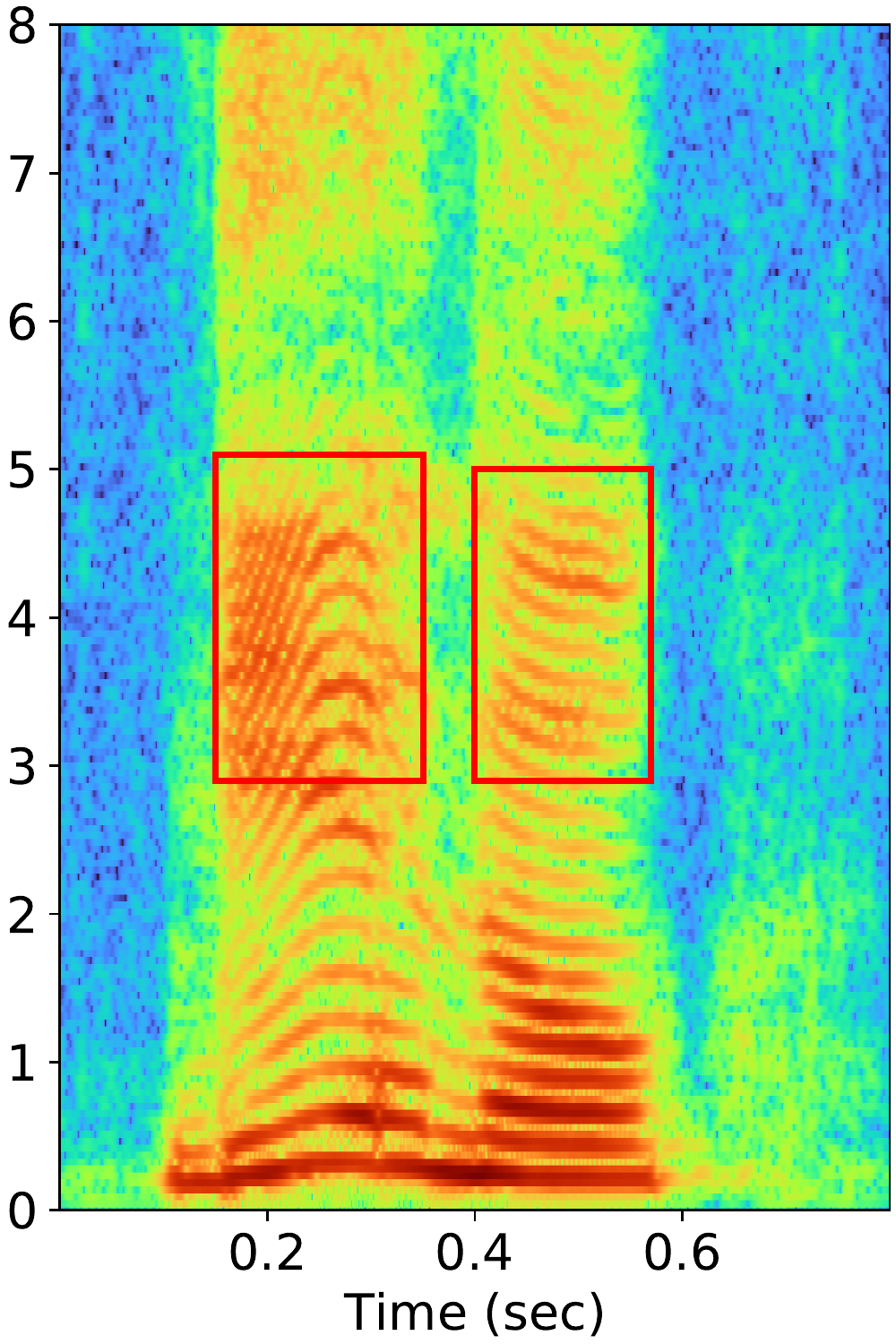,width=28mm}}
	\centerline{(c)}  \medskip
	\end{minipage}	
	\caption{\small Spectrograms of (a) natural speech, (b) generated speech from the conventional Parallel WaveGAN (S2), and (c) generated speech from the proposed Parallel WaveGAN (S7). \rrredit{As demonstrated in rectangle areas, our proposed method is able to model spectral harmonics more accurately.}}	
	\vspace*{-2mm} 
	\label{fig:spec_anasyn}
	\end{figure}
	\begin{table*}[!t]   
	\begin{center}         
	\caption{\small MOS \eeeeedit{test} results with 95\% confidence intervals: Acoustic features generated from the FastSpeech 2 acoustic model were used to compose the input auxiliary features.}  
    \vspace{1mm}
	\label{tab:mos_tts}
	\scalebox{0.99}{
	{\small        
	\begin{tabular}{clcccc}
	\Xhline{2\arrayrulewidth}
    \multirow{2}*{System} & \multirow{2}*{{~~~~~~~~Model}} & \multicolumn{4}{c}{MOS} \\
    & & F1 & F2 & M1 & M2 \\
    \hline
    S1 & FastSpeech 2 + WaveNet      & 3.90$\pm$0.11  & 3.81$\pm$0.10  & 3.43$\pm$0.11  & 3.09$\pm$0.10 \\
    S2 & FastSpeech 2 + PWG          & 3.76$\pm$0.11  & 3.62$\pm$0.11  & 3.63$\pm$0.11  & 3.78$\pm$0.10 \\
    S3 & FastSpeech 2 + PWG-cGAN-D & 4.02$\pm$0.10  & 4.03$\pm$0.10  & 4.16$\pm$0.10  & 4.06$\pm$0.10  \\
    \textbf{S7} & \textbf{FastSpeech 2 + PWG-V/UV-D (proposed)} & \textbf{4.20$\pm$0.10} & \textbf{4.18$\pm$0.09}  & \textbf{4.21$\pm$0.09} & \textbf{4.31$\pm$0.09} \\
    \bottomrule
    R1 & Recordings & 4.63$\pm$0.08  & 4.67$\pm$0.07  & 4.61$\pm$0.08  & 4.64$\pm$0.08 \\
    \Xhline{2\arrayrulewidth}
	\end{tabular}}    
	}
	\end{center} 
    \vspace{-2mm}
	\end{table*}

	We performed mean opinion score (MOS)\footnote{Audio samples are available at the following URL:\\\url{https://r9y9.github.io/demos/projects/icassp2021/}} tests to investigate the effectiveness of our proposed method.
    \redit{Seventeen} native Japanese speakers were asked to make quality judgments about the synthesized speech samples using the following five possible responses: 1 = Bad; 2 = Poor; 3 = Fair; 4 = Good; and 5 = Excellent. 
    In total, 20 utterances were randomly selected from the evaluation set and were then synthesized using the different models. 

    Table~\ref{tab:mos_anasyn} shows the MOS test results with respect to different neural vocoders.
    The findings can be analyzed as follows.
    (1) Among all speakers, the Parallel WaveGAN system using the conditional discriminator (S3) obtained a better score than the baseline \rredit{WaveNet (S1) and Parallel WaveGAN (S2)} systems.
    The results confirmed the effectiveness of incorporating acoustic features into the discriminator through conditional information.
    (2) The systems using poorly configured voicing-aware discriminators (S4, S5, and S6) performed worse than S3. More specifically, they performed even worse than the baseline Parallel WaveGAN (S2) in some cases (e.g., comparing S2 and S5). 
    Notably, the synthetic male voices contained buzzy noise, which resulted in significantly lower scores than the baseline Parallel WaveGAN.
    (3) Finally, the system with the proposed voicing-aware discriminators (S7) obtained the best scores and consistently outperformed the \rredit{other systems (from S1 to S6)}.
    The results proved the importance of the discriminator design and the effectiveness of our proposed approach.
    The benefits of our method can also be confirmed in spectrogram visualization; as shown in Fig.\ref{fig:spec_anasyn}, 
    the proposed method was able to better reconstruct the spectral harmonics.

    \subsection{Text-to-speech}

    To further verify the effectiveness of the proposed method \rredit{within the TTS framework}, we combined the proposed Parallel WaveGAN with a FastSpeech 2 based acoustic model~\cite{ren2020fastspeech}.
    This model was configured based on the setup of FastSpeech 2, except for a few modifications as follows:
	we changed the variance predictor module to operate on phoneme-level rather than frame-level~\cite{lancucki2021fastpitch}. 
	Manually annotated phoneme alignment was used instead of performing forced alignment. 
	The model used accent \rredit{information} as an external input to better model pitch accents of Japanese~\cite{yasuda2019investigation}.

    For evaluation, we trained four speaker-dependent acoustic models.
    At the training stage, a dynamic batch size with an average of 24 samples was used for making a mini-batch~\cite{li2019close,hayashi2020espnet}, and the models were trained for 200K iterations.
	In the synthesis step, the input phoneme and accent sequences were converted to the corresponding acoustic parameters by the FastSpeech 2 model. 
	By inputting the resulting acoustic parameters, the vocoder models generated the time-domain speech signals.

	To evaluate the quality of the generated speech samples, we performed naturalness MOS tests.
    The test setups were the same as those described in section~\ref{sec:evaluation}, except that we excluded the systems with poorly designed discriminators (S4, S5, and S6 in Table~\ref{tab:mos_anasyn}, respectively)
    
    The results of the MOS tests are shown in Table~\ref{tab:mos_tts}.
    \rredit{Similar to the analysis/synthesis results}, the proposed system with voicing-aware conditional discriminators (S7) achieved the best scores among all speakers.
    In particular, the proposed method significantly outperformed the baseline \rredit{WaveNet \rredit{(S1)} and Parallel WaveGAN \rredit{(S2)}} systems\rredit{, and even} the improved Parallel WaveGAN with conditional discriminator (S3). 
    \redit{Note that the MOS of the TTS samples tended to be higher than that of analysis/synthesis samples. This \rredit{result} was because the unwanted artifacts produced by the analysis/synthesis process were statistically excluded during the generation process. Most listeners preferred consistent results of the predicted duration and accent than \redit{those of} the ground-truth.}

\section{Conclusion}
\label{sec:summary}
    
    We proposed voicing-aware conditional discriminators for a Parallel WaveGAN-based TTS system.
    Our framework \rredit{incorporated} a projection-based conditioning method into the discriminator and \rredit{divided} it into two separate discriminators.
    By controlling the voiced and unvoiced speech segments independently, the performance of each discriminator was significantly improved, which allowed the generator to produce more \rredit{natural} speech waveforms.
    The experimental results demonstrated the superiority of our proposed method over the conventional Parallel WaveGAN and similarly configured WaveNet systems.
    Future work includes improving the \eeeedit{performance of} \rredit{the} generator by utilizing the voicing information of speech.

\section{Acknowledgements}

    This work was supported by Clova Voice, NAVER Corp., Seongnam, Korea.

\bibliographystyle{IEEEbib-abbrev}
\bibliography{refs}
\end{document}